\documentclass[aps,prl,twocolumn,showpacs, superscriptaddress]{revtex4-1}

\usepackage{graphicx}
\usepackage{color}
\usepackage{soul}

\begin{document}


\title{Non-Markovianity hinders Quantum Darwinism}

\author{Fernando Galve} 
\email[]{fernando@ifisc.uib-csic.es} 
\author{Roberta Zambrini} 
\email[]{roberta@ifisc.uib-csic.es}
\affiliation{IFISC (UIB-CSIC), Institute of Cross-Disciplinary Physics and Complex Systems, UIB Campus, 07122 Palma de Mallorca, Spain}

\author{Sabrina Maniscalco} 
\email[]{smanis@utu.fi} \homepage[]{www.openquantum.co.uk}
\affiliation{Turku Center for Quantum Physics, Department of Physics and
Astronomy, University of Turku, FIN-20014 Turku, Finland}

\date{\today}

\begin{abstract}
In this Letter we investigate Quantum Darwinism and the emergence of a classical
world from the quantum one in connection with the spectral properties of the
environment. We use a microscopic model of quantum environment in which, by
changing a simple system parameter, we can modify the information back flow from
environment into the system, and therefore its non-Markovian character. We show
that the presence of memory effects hinders the emergence of classical objective
reality. We conjecture that this phenomenon is connected to the absence of a
measurement-scheme interpretation for quantum trajectories in the non-Markovian
regime.

\end{abstract}


\maketitle

{\sl Introduction-} Quantum Darwinism is a fascinating theory that explains the
emergence of a classical objective reality in terms of proliferation of
information about certain states of a quantum system into the environment \cite{zurekNAT,brandao}.
We live in a quantum Universe, the behaviour of all microscopic constituents
being described by the laws of quantum physics. There is overwhelming evidence
that this incredibly successful theory applies at all scales. Why then the
macroscopic objects populating our everyday reality are only found in a much
smaller subset of states, consistent with classical laws? 

The emergence of the classical objective world from the quantum one has been
debated and investigated since over a century \cite{Exploring}. 
A significant advance in our understanding of the quantum measurement problem
was achieved through the theory of environment induced decoherence \cite{deco1}, showing how
continuous monitoring by the environment destroys fragile quantum information
and allows only certain classical states to survive \cite{deco2}. Quantum Darwinism takes a
further step promoting the role of the environment from a passive sink of
coherence to the active carrier of information about the system. Indeed, as
Zurek observes \cite{zurekNAT}, generally we do not measure quantum systems directly,
rather we infer their properties by the observation of parts of the environment.
The question addressed by quantum Darwinism is how certain states of the system
produce multiple redundant imprints in uncorrelated fractions of the
environment. Such fractions can then be measured by many independent observers
that will detect the same property of the system, without perturbing it. This
illustrates the emergence of an objective classical reality from the quantum
probabilistic world. A recent breakthrough in the field \cite{brandao}
has shown that spreading of the classical information about a pointer observable
into the environment is a generic feature of quantum mechanics. Furthermore it
has been recently shown that for all models of pure decoherence \cite{zurekCHER} this spreading leads
to agreement of an objective classical reality among different observers. The same has been
shown to be true for the photonic environment \cite{zurekPHOTON,korbicz}, and in general for 
a two level system under a controlled unitary evolution for the bath \cite{horodecki}. Experimental 
evidence for its workings in a quantum dot scenario has been given in \cite{dots}.

The main result of this Letter is the discovery of a strong correlation between the 
non-Markovian character of the open quantum system dynamics, indicated by the presence of
information flowback \cite{nonMarkov}, and the inhibition of redundancy and hence objectivity. 
In the non-Markovian dynamics setting \cite{nonMarkov,nonMarkov2,resource}, indeed, measurements by the observers would inevitably subtract part of the information
present in the environment that could feedback into the system, hence perturbing
the system's subsequent state.

{\sl Model-}
We illustrate this effect by looking at a paradigmatic model, the quantum Brownian
oscillator. This model has been extensively studied in the literature (see, e.g., Ref. \cite{deco2}) to describe vibrational/bosonic modes dissipating into bosonic reservoirs.
Its behavior in terms of quantum Darwinism has been studied in the case of
Markovian evolution \cite{zurekOSC}. The non-Markovian properties of this model for non-Ohmic
spectral densities were explored recently in \cite{vasile}, where the relation between
resonant/detuned situations and non-Markovianity was pinpointed. We here fill the gap and show that non-Markovian behavior leads to breaking of the process of quantum Darwinism, thus
hindering the production of stable records in the bath.

The model consists of a quantum harmonic
oscillator of frequency $\omega_S$ coupled through the bilinear interaction
$-\kappa x_S\sum_{k=1}^N c_kx_k$ to a quantum environment of $N$ harmonic
oscillators having frequencies $\omega_k=\omega_0+k\Delta$, with
$\Delta=(\omega_R-\omega_0)/N$, and $x_S$ and $x_k$ position operators of system
and environment oscillators, respectively. The spectral
density is given by
\begin{equation}
J(\omega)=\sum_{k=1}^N\frac{c_k^2}{\omega_k}\delta(\omega-\omega_k),
\end{equation}
 which becomes a function in the continuum limit. In typical scenarios an Ohmic ($s=1$) frequency dependence
$J(\omega)\sim\omega^sf(\omega/\omega_R)$ with $f(\omega/\omega_R)$ a frequency cutoff function which decays fast
for $\omega>\omega_R$ is used, with most variations covering the $s\neq 1$ cases.
A possible microscopic derivation of such Ohmic dissipation is based on the Rubin model \cite{Rubin}, consisting
of a homogeneous linear harmonically coupled chain of equal on-site frequency ($\omega_0$) oscillators with coupling strength $g$, leading
to a spectral density $J(\omega)=\kappa\sqrt{\omega^2-\omega_0^2}\sqrt{\omega_R^2-\omega^2}$ with $\omega_R=\sqrt{\omega_0^2+4g}$. This
microscopic model reproduces an Ohmic spectral density for $\omega_0=0$, while it allows for the study
of non-Markovian dynamics when the system is detuned with respect to such bath.


We consider the case in which the system oscillator is initially prepared in a
momentum-squeezed state with squeezing parameter $r$ and the environment is in the vacuum
state.  Because the system and bath couple through position, this is going to be
the classically `recorded' observable of the system, meaning that the initial position
spread will be redundantly stored in fragments of the environment. In this sense we expect, if
perfect quantum Darwinism occurs,
the initial state of the system $|\psi(0)\rangle_S\propto\int dx \exp{(-x^2/2\sigma^2)}|x\rangle$,
with $2\sigma^2=\exp{(2r)}$, to be perfectly broadcasted \cite{brandao,korbicz} into the environment
$|\psi\rangle_{SB}\propto \int dx\exp{(-x^2/2\sigma^2)}|x\rangle\bigotimes_i|\psi_i(x)\rangle$
as independent, perfectly distinguishable ($\langle \psi_i(x)|\psi_i(x')\rangle\sim\delta_{x,x'}$),
branches which are conditional on each initial position. It is interesting to note that, using the
state analysis in \cite{zurekOSC}, this ideal situation can be shown to match  a `measure and prepare'
map \cite{brandao} for system+fraction and furthermore that such branching state can be written in `spectrum broadcast' form \cite{korbicz}
\begin{equation}
\rho_{S,f}=\int dx\, p(x)|x\rangle\langle x|\otimes\rho_{i_1}(x)\otimes\rho_{i_2}(x)\otimes ...\otimes\rho_{i_f}(x)
\end{equation}
with $\rho_{i_n}(x)=|\psi_{i_n}(x)\rangle\langle\psi_{i_n}(x)|$, $\rho_{i_n}(x)\rho_{i_n}(x')=\langle \psi_{i_n}(x)|\psi_{i_n}(x')\rangle\, |\psi_{i_n}(x)\rangle\langle\psi_{i_n}(x')|=0$ for $x\neq x'$,
where the fraction is composed by oscillators $\vec{i}=\{i_1,i_2,...i_f\}$.
In fact \cite{zurekOSC} showed that $\langle \psi_{i_n}(x)|\psi_{i_n}(x')\rangle=\exp[-d_{i_n}(t)(x-x')^2]$, where $d_{i_n}(t)$ is 
a decohering factor highlighting the strength with which fraction ${i_n}$ of the bath is able to decohere 
the system. This means that perfect Darwinism is reached when such factor is big enough.
Intuitively, if this ideal situation is achieved, each fragment of the 
environment can perfectly distinguish the different classical position records of the system, and just by
interrogating a fraction ({\it any} fraction, of {\it any} size) we can gain the information on the label $x$.


Our main goal is to discuss the connection between non-Markovianity and the
dynamical onset of objectification. We will therefore use and compare the
non-Markovianity measure for continuous variables systems introduced in Ref.
\cite{NM} on the one hand, and the mutual information between system and fractions
of the environment \cite{zurekOSC} on the other hand.

We detect and quantify the presence of memory effects in terms of information
flow back  as indicated by the non-monotonic behaviour of fidelity between pairs
of states $F(\rho_1,\rho_2)$. Fidelity monotonically increases under the action
of completely positive and trace preserving maps, indicating a decrease in state
distinguishability and hence a loss of information on the system due to the
action of the environment. A temporary decrease of fidelity for certain time
intervals therefore signals a partial increase of information on the system,
i.e., information flow back. The corresponding measure of non-Markovianity is
obtained by optimizing over all pairs of states \cite{NM}
\begin{equation}
{\cal N}={\rm max}_{\rho_1, \rho_2} \int_{dF/dt < 0} 
\frac{dF(\rho_1,\rho_2)}{dt} dt, \label{NMMeas}
\end{equation}
where the integral extends over all time intervals in which the derivative of
the fidelity is negative.
In practice, we restrict our attention to Gaussian states, and calculate the
corresponding Gaussian measure of non-Markovianity.

The open quantum system model here considered has the property that, by varying
the frequency of the system oscillator with respect to the spectral
distribution, we change the non-Markovian character of the open system dynamics
\cite{vasile}.
Specifically, higher values of non-Markovianity are obtained close to the
edges of the spectral distribution while on resonance non-Markovianity is
minimal, as we further illustrate in the following.

\begin{figure}
\includegraphics[width=\columnwidth]{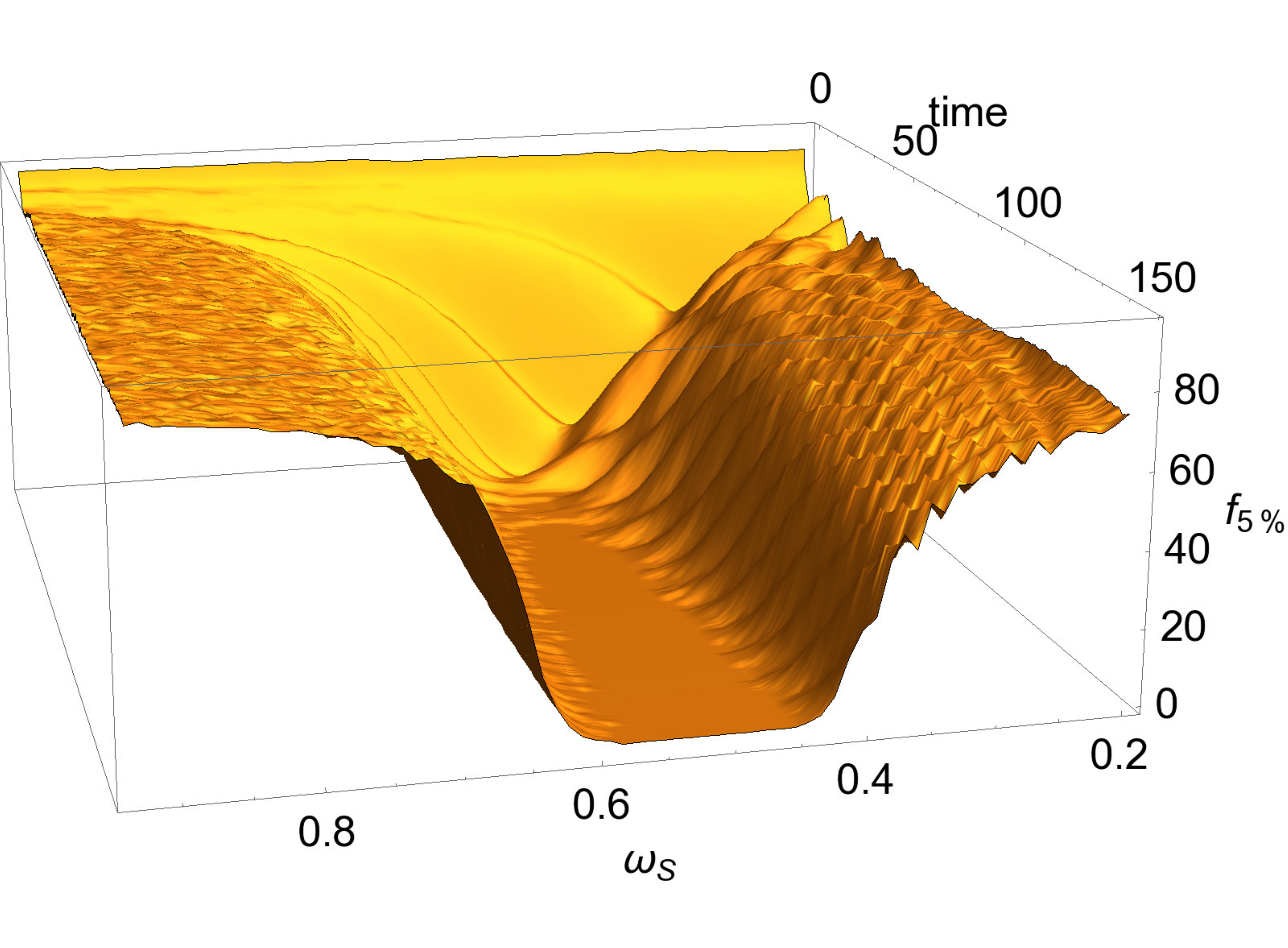}
\caption{(Colors online) Behaviour of $f_{\delta}=1/R_{\delta}$ for $\delta=
5\%$ as a function of time (a.u.) and of the system frequency $\omega_S$ for a
bath of $N=300$ oscillators with central frequency $\omega_0=0.3$ and cutoff frequency
$\omega_R=0.7$. The initial squeezing parameter is $r=10$.}
\label{fig1}
\end{figure}

\begin{figure}
\includegraphics[width=\columnwidth]{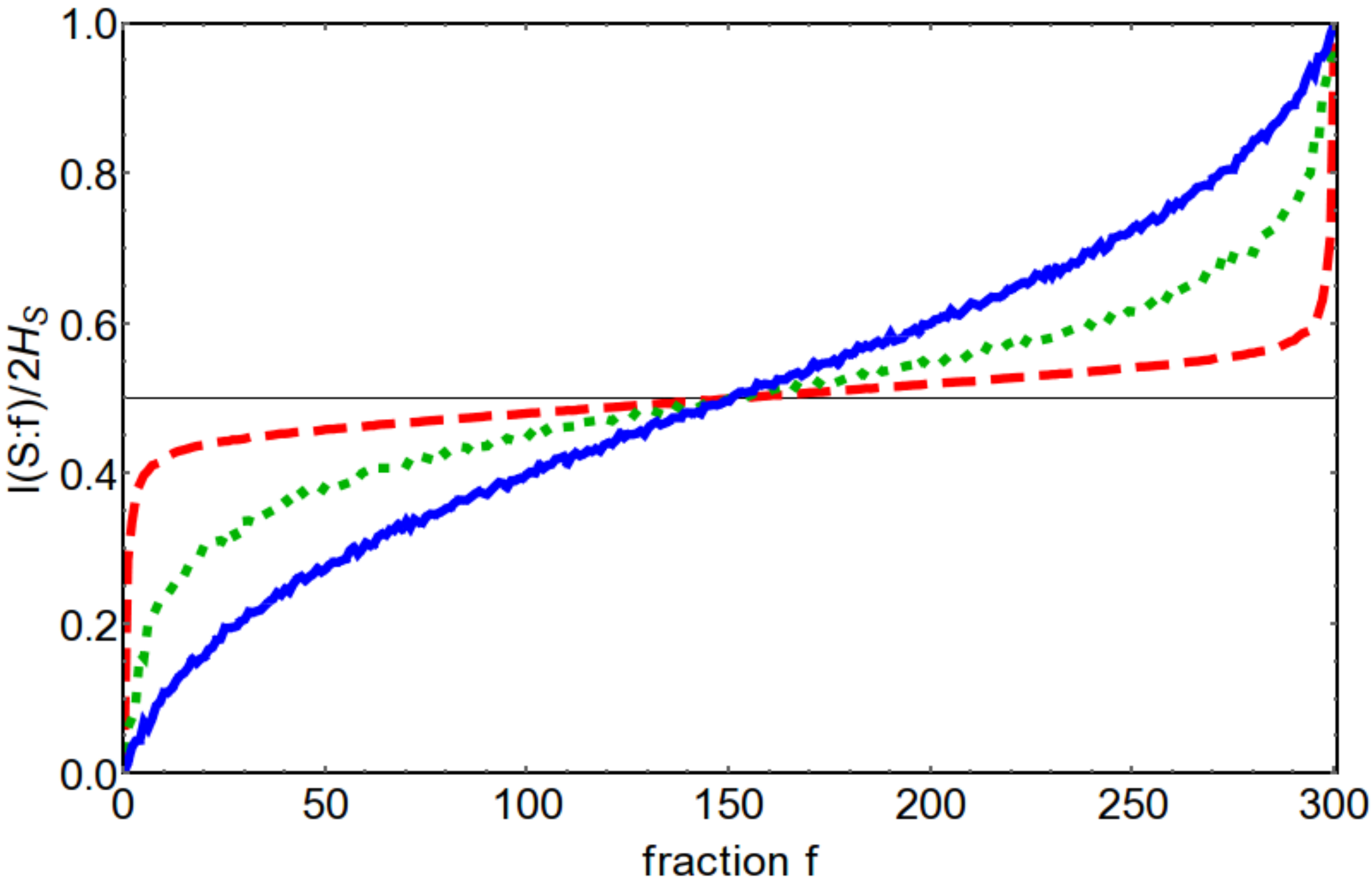}
\caption{(Colors online) Mutual Information $I(S:f)/ H_S$ as a function of the
fraction $f$ for $t=40$ (a.u.) and $\omega_S=0.3$ (red/dashed line), $\omega_S=0.7$
(green/dotted line) and $\omega_S=1$ (blue/solid line), with $r=3$. All other parameters are the same as
in Fig. 1.}
\label{fig2}
\end{figure}

{\sl Results-}
In describing quantum Darwinism the key quantity of interest describes
correlations between the system and fragments of the environment of size $f$.
These correlations are given by the mutual information 
\begin{equation}
I(S:f) = H_S + H_f -H_{S,f},
\end{equation}
with $H_S$ and $H_f$ the individual von Neumann entropies of system and
environmental fraction, and $H_{S,f}$ the entropy of the combined system. If the
initial state is pure the quantity $H_S$ is the entropy of the system due to
decoherence and it therefore measures all the information about $S$ available
from either the system or the environment. The emergence of an objective reality
through decoherence is indicated by the existence of a plateau around $f=1/2$ in
the plot of $I(S:f)$ as a function of $f$. More precisely the values of $I(S:f)$
quickly raise to $H_S$, indicating that already small factions of the
environment contain almost all information classically accessible on the system.
Redundancy $R_{\delta}=1/f_{\delta}$ is the number of independent fragments of
the environment that supply almost all classical information on the system, i.e.
$(1-\delta) H_S$. Large redundancy, or equivalently small values of
$f_{\delta}$, show that the state of the system can be found independently by
many observers  (probing e.g. independent fractions of different sizes), 
who agree on the outcome, by measuring parts of the environment without
disturbing the system.  An important point is also the fact that an observer needs
not collect information on a huge set of environmental degrees of freedom, impossible in practical
terms, in order to get an idea of the state of the system.

We choose randomly fractions of the environment and investigate the dynamical
onset of objectification for different values of the system frequency
$\omega_S$. As an example, we plot in Fig. 1 the behaviour of $f_{5\%}$. The
plot clearly shows that quantum Darwinism, indicated by small values of
$f_{5\%}$ occurs very rapidly for values of the system frequency resonant with
the peak of the spectral density while it is strongly inhibited in the
off-resonant regime. This result is confirmed by the plots of Fig. 2 where we
show the the asymptotic mutual information as a function of $f$ for resonant
system frequency (red curve), at the edge of the spectral distribution (green
curve) and for off-resonant frequency (blue curve). It is evident that the
plateau disappears when the system frequency moves away from the peak of the
spectrum. 

\begin{figure}
\includegraphics[width=\columnwidth]{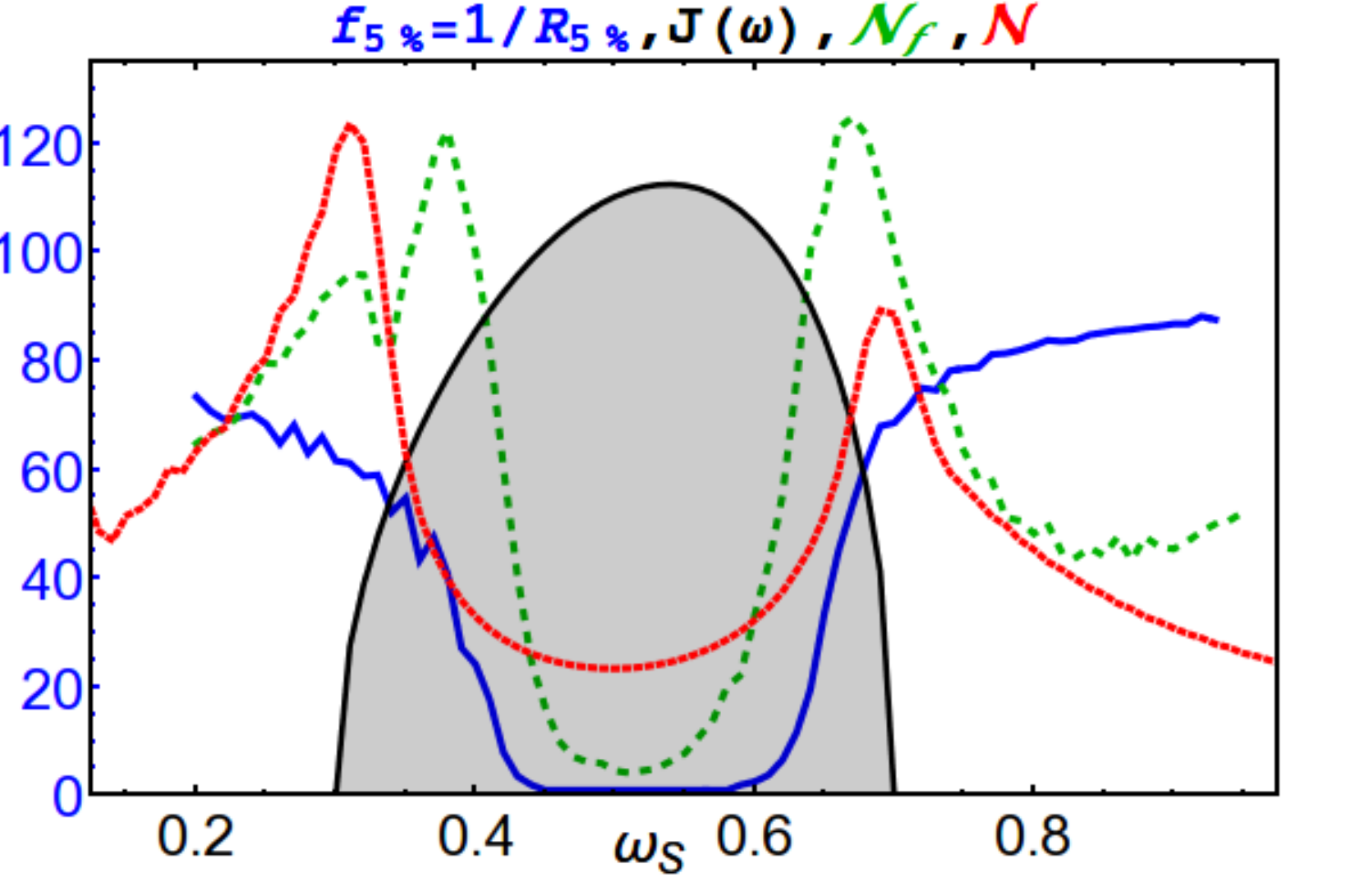}
\caption{(Colors online) Spectral density $J(\omega)$ (shaded gray),
non-Markovianity ${\cal N}$ (red/dotted line), fraction
$f_{\delta}=1/R_{\delta}$ (blue/solid line) with $\delta = 5\%$, and
non-monotonicity of $f_\delta$ ${\cal N}_f$ (green/dashed line) for
$t=150$, initial squeezing $r=10$. Other parameters as in Fig. 1. All plots are in a.u.
}
\label{fig3}
\end{figure}

In order to link this behaviour with memory effects arising from the structural
features of the frequency spectrum we calculate the non-Markovianity measure
${\cal N}$ of Eq. (\ref{NMMeas}) and compare its behaviour  with the behaviour
of the fraction $f_{5\%}$ when varying $\omega_S$. As one can see from Fig. 3,
information flow back, and hence ${\cal N}$,  is maximal at the edge of the
spectrum and minimal at the centre. This is consistent with the fact that on
resonance the system energy is rapidly dissipated into the environment. As we
move towards the edges of the distribution, the system oscillator is more
strongly sensitive to the variations in the form of the spectrum and memory
effects become dominant. When off-resonance, oscillatory behaviour dominates the
dynamics. Small amplitude oscillations are also present in the redundancy (See
Fig. 1) and in the dynamics of the mutual information $I(S:f)$. We quantify this
non-monotonic behaviour of $f_\delta$ by introducing its non-monotonicity
defined in a similar fashion to ${\cal N}$, namely
 \begin{equation}
 {\cal N}_f= \int_{df/dt > 0}  \frac{df}{dt} dt.
 \end{equation} 
 This quantity is plotted in Fig. 3 (green dashed line) and it shows a good
qualitative agreement with the behaviour of the non-Markovianity measure ${\cal
N}$.
There are hence three main regimes where we can highlight the 
connection between non-Markovianity and quantum Darwinism: when in resonance,
the system dissipates monotonously into the bath, meaning a Markovian evolution
and a constant tendency towards a perfect quantum Darwinism situation where 
stable redundant records are established in the environmental fractions. Out of resonance
we have exactly the opposite situation, practically no dissipation and no bath's
information gain about the system (very poor Darwinism). However, near the frequency
edges, information and energy flow back and forth from the system, meaning non-monotonic 
dissipation and non-Markovianity as well as non-monotonic (${\cal N}_f>0$) creation
of records leading to poor Darwinism (in the sense of higher $f_{5\%}$). 
The very definition of record requires temporal stability, being this the main motivation
to introduce the indicator ${\cal N}_f$.
All
these cases can also be understood in terms of the decoherence factor $d_i(t)$ appearing
in the system+fraction state, which has spectrum-broadcast form, that we have discussed
before. If the decoherence factor, for all $i$, grows in time to big values (resonant case),
we will have perfectly distinguishable records in each fragment and perfect Darwinism,
as in \cite{zurekOSC}. Out of resonance, $d_i(t)$ do not grow and records are fuzzy while in the vicinity of band edges,
$d_i(t)$ are oscillating.

{\sl Conclusions-}
We have studied the connection between non-Markovian dynamics, characterised by
information flow back, and the emergence of an objective classical reality
through the proliferation of classical copies of the system state in the normal
modes of the environment. We have shown that, by modifying the properties of the
spectral density, and, in particular, by changing the resonance condition
between the system oscillator and the reservoir spectral peak, we can inhibit
quantum Darwinism by modifying the Markovian character of the dynamical map.

The quantum Darwinism scenario is deeply connected to a well known description of
the dynamics of open quantum systems in terms of quantum trajectories. It is
well known that Markovian open quantum systems described by master equations of
Lindbad-Gorini-Kossakowski-Sudarshan form \cite{LGKS} can be described by means of a
stochastic Schr\"odinger equation. Quantum trajectories are pure state solutions
of such an equation and can be interpreted as evolution of a single quantum
system while its environment is continuously monitored. The deterministic
evolution of the state vector is interrupted by the occurrence of quantum jumps
whose statistics (waiting time distribution) is linked to the decay rates of the
master equation. The evolution of the reduced density matrix of the system is
obtained as an ensemble average over single trajectories. 

A longstanding open problem of open quantum system theory is the existence of a
physical interpretation for single quantum trajectories in the non-Markovian
regime, i.e. when memory effects due to long-lasting correlations between system
and environment occur. While a definitive answer has not yet been given, it is believed that non-Markovian
trajectories do not have a measurement scheme interpretation \cite{Gambetta}. Indeed, contrarily
to the Markovian case, measuring the environment would subtract information that
may feedback into the system at a later time, as a result of reservoir memory,
hence modifying the dynamics of the system. This is clearly illustrated by the non-Markovian
quantum jumps unravelling \cite{NMQJ}.
 
We believe that the very same reasons for which genuine non-Markovian quantum trajectories do not seem to exist 
prevents observers to measure part  of the environment, in order to infer
information on the system, without disturbing the system. In other words, because the production of stable and redundant copies of classical
information about the system is linked with monotonous dissipation into the bath,
any flow of information back into the system can be directly interpreted as diminishing
these records, whence the relation between non-Markovian behavior and non-monotonic
redundant registers formation.

Besides the fundamental interest, our results give a new perspective on the active debate on whether or not
non-Markovianity can generally lead to more efficient quantum technologies. A
number of investigations suggest that this might actually be the case in certain
specific scenarios \cite{resource} but a formal answer in the framework of abstract
resource theory is still missing. Our study supports the conjecture that
non-Markovian dynamics might be useful for quantum technologies showing that
memory effects push the boundary between the quantum and the classical world,
inhibiting the emergence of classicality. In this sense reservoir engineering
techniques may open a
new and yet unexplored door into  the quantum realm.

{\sl Acknowledgements-}
Funding from EU (FEDER), CSIC (JAE contract), UIB (visiting professors program), and MINECO (TIQS project) is acknowledged.

\end{document}